\documentclass[twocolumn,superscriptaddress,showpacs,aps,prl,longbibliography]{revtex4-2}
\usepackage{graphicx} 
\usepackage{dcolumn}
\usepackage{bm}
\usepackage{graphicx}
\usepackage{times}
\usepackage{soul}
\usepackage{color}
\usepackage{braket}
\usepackage{natbib}
\usepackage{floatrow}
\usepackage{physics}
\usepackage{comment}
\usepackage{amssymb}

\usepackage{soul,xcolor}
\usepackage[normalem]{ulem}
\definecolor{black}{rgb}{0,0,0}
\definecolor{blue}{rgb}{0,0,1}
\definecolor{green}{rgb}{0,1,0}
\definecolor{red}{rgb}{1,0,0}
\definecolor{brown}{rgb}{0.4,0.2,0}
\definecolor{darkgreen}{rgb}{0,0.7,0}
\definecolor{darkblue}{rgb}{0.0,0.0,0.5}
\definecolor{red}{rgb}{1,0,0}
\definecolor{deepmagenta}{rgb}{0.8, 0.0, 0.8}

\usepackage[colorlinks,
citecolor = blue,
linkcolor = blue,
urlcolor  = blue,
anchorcolor = blue,
breaklinks = true
]{hyperref}

\usepackage{tikz, mathtools}
\DeclarePairedDelimiter{\BRA}{\langle}{\rvert}
\DeclarePairedDelimiter{\KET}{\lvert}{\rangle}
\tikzset{dot/.style={draw, thin, circle, fill, outer sep=0pt, inner sep=0pt, minimum width=1mm}}

\newcommand{\plaqx}{\tikz[baseline=.5ex, scale=.4]{
		\draw(0,0)node[dot](A){}--(60:1)node[dot](B){}(1,0)node[dot](D){}--++(60:1)node[dot](C){};
		\draw[line width=1.5pt](A)--(D)(B)--(C);
	}
}
\newcommand{\plaqy}{\tikz[baseline=.5ex, scale=.4]{
		\draw[line width=1.5pt](0,0)node[dot](A){}--(60:1)node[dot](B){}(1,0)node[dot](D){}--++(60:1)node[dot](C){};
		\draw(A)--(D)(B)--(C);
	}
}

\newcommand{\recx}{\tikz[baseline=.5ex, scale=.4]{
		\draw(0,0)node[dot](A){}--(0,1)node[dot](B){}(1,0)node[dot](D){}--++(0,1)node[dot](C){};
		\draw[line width=1.5pt](A)--(D)(B)--(C);
	}
}
\newcommand{\recy}{\tikz[baseline=.5ex, scale=.4]{
		\draw[line width=1.5pt](0,0)node[dot](A){}--(0,1)node[dot](B){}(1,0)node[dot](D){}--++(0,1)node[dot](C){};
		\draw(A)--(D)(B)--(C);
	}
}

\graphicspath{{./Figures/}}

\def \IBK{Institute for Theoretical Physics, University of Innsbruck, Innsbruck A-6020, Austria}
\def \IQOQI{Institute for Quantum Optics and Quantum Information,
	Austrian Academy of Sciences, Innsbruck A-6020, Austria}
\def \planqc{planqc GmbH, Garching D-85748, Germany}

\begin{document}

\title{Quantum dimer models with Rydberg gadgets}
	\author{Zhongda Zeng}\affiliation{\IBK}\affiliation{\IQOQI}
	\author{Giuliano Giudici}\affiliation{\IBK}\affiliation{\IQOQI}\affiliation{\planqc}
	\author{Hannes Pichler}\affiliation{\IBK}\affiliation{\IQOQI}
\preprint{}

\begin{abstract}
The Rydberg blockade mechanism is an important ingredient in quantum simulators based on neutral atom arrays. It enables the emergence of a rich variety of quantum phases of matter, such as topological spin liquids. The typically isotropic nature of the blockade effect, however, restricts the range of natively accessible models and quantum states. In this work, we propose a method to systematically overcome this limitation, by developing gadgets, i.e., specific arrangements of atoms, that transform the underlying Rydberg blockade into more general constraints. We apply this technique to realize dimer models on square and triangular geometries. In these setups, we study the role of the quantum fluctuations induced by a coherent drive of the atoms and find signatures of $U(1)$ and $\mathbb{Z}_2$ quantum spin liquid states in the respective ground states. Finally, we show that these states can be dynamically prepared with high fidelity, paving the way for the quantum simulation of a broader class of constrained models and topological matter in experiments with Rydberg atom arrays. 
\end{abstract}

\date{\today}

\maketitle

\paragraph{Introduction. --} \label{sec:introduction}

\begin{figure}[t]
    \centering  
    \includegraphics[width=1.0\columnwidth]{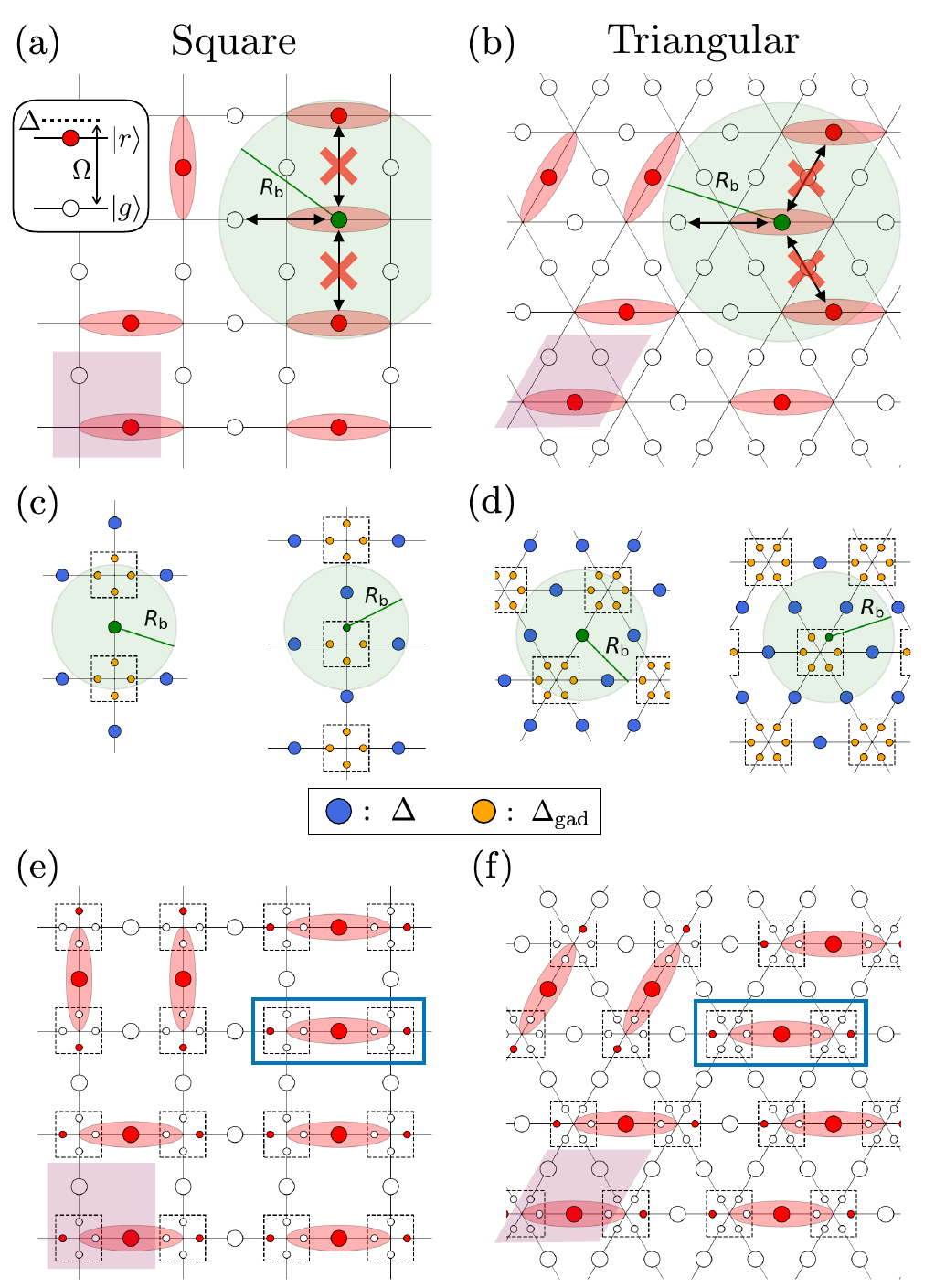}
    \caption{ 
        \textbf{Rydberg gadgets for the dimer constraint.}
        Left and right panels correspond to the square and triangular lattices, respectively. (a,b) Typical dimer fully-packed configurations. Red (white) dots represent atoms in the Rydberg (ground) state (see the inset of (a)). The red crosses indicate the undesired blockade for the selected blockade radius $R_{b}$.
        (c,d) Illustration of the Rydberg gadgets. Gadgets are placed on the vertices of the lattice and consist of the orange atoms. The green disks indicate the blockade radius required to realize the dimer constraint.
        (e,f) Rydberg configurations in the gadget model corresponding to the dimer coverings in (a,d). Each dimer consists of one excited atom on the edge and two excited atoms on the gadgets at the extrema (blue rectangle). The purple-shaded square is the lattice unit cell.
        }
    \label{fig:illustration_encoding}
\end{figure}
Preparing and studying strongly correlated many-body quantum states is one of the central goals of near-term quantum devices~\cite{Preskill2018quantumcomputingin}. Quantum spin liquids (QSLs) are a particularly interesting class of such states, that feature many exotic properties, including long-range entanglement, anyonic excitations, and topological ground state degeneracy~\cite{Savary_2017,RevModPhys.89.041004,Sachdev_2019,balents_spin_2010}. 
In addition, QSLs offer intriguing possibilities for applications in quantum information processing, where they could form the basis of effective error-correcting schemes for robust quantum memories~\cite{kitaev1}, or for topological quantum computation~\cite{KITAEV20032,KITAEV20062,RevModPhys.80.1083}. 	
Recently, remarkable progress has been achieved in realizing some of these topologically ordered states with several quantum simulator platforms, using a variety of state preparation techniques. 
In digital approaches, the target states are prepared by applying circuits consisting of entangling gates to simple product states~\cite{toric_google,iqbal2023topological,iqbal2023creation}. In contrast, in analog approaches, dynamical preparation protocols with a controllable, time-dependent Hamiltonian are employed, e.g. using the adiabatic principle~\cite{doi:10.1126/science.abi8794,Giuliano_PhysRevLett.129.090401,Ruben_sahay2023quantum}. The latter has the advantage that the prepared states can potentially be stabilized by a Hamiltonian and dynamical signatures of QSL phases could be probed. 
 
Experimentally, the analog approach to realizing QSLs in synthetic quantum matter has been pioneered with Rydberg atom arrays~\cite{doi:10.1126/science.abi8794}. These platforms enable the realization of highly programmable spin models, that have been explored for quantum simulation~\cite{bernien_probing_2017,Labuhn_2016,keesling_quantum_2019,ebadi_quantum_2021,doi:10.1126/science.abg2530,doi:10.1126/science.abi8794}, quantum optimization~\cite{doi:10.1126/science.abo6587}, and quantum computation~\cite{Levine_2019,Bluvstein_2022,evered_high-fidelity_2023,bluvstein_logical_2023}. In these setups, atoms are trapped in optical tweezers or lattices and internal states are employed to represent spin degrees of freedom~\cite{browaeys_many-body_2020}. Control is enabled by coherent laser drive and the geometric arrangement of the atoms~\cite{doi:10.1126/science.aah3752,doi:10.1126/science.aah3778}. A crucial ingredient is the so-called blockade effect~\cite{PhysRevLett.85.2208,PhysRevLett.87.037901,gaetan_observation_2009}, which prevents two atoms from being simultaneously excited to Rydberg states when their distance is smaller than the blockade radius. 
This mechanism, when combined with specific geometries of the array, can be used to introduce constraints on the allowed spin configurations, e.g., to mimic the Gauss's law of lattice gauge theories~\cite{PhysRevX.10.021041,PhysRevLett.129.195301,PhysRevX.10.021057,Homeier2023}. Moreover, it is a key ingredient for providing the geometric frustration that leads to the emergence of QSLs from many-body Rydberg Hamiltonians~\cite{PhysRevX.11.031005,doi:10.1073/pnas.2015785118,PhysRevLett.130.043601,PhysRevX.12.041029,QSLtrimer,quera2022trimer}.

The Rydberg blockade is typically isotropic, strongly restricting the class of non-trivial constraints that can be implemented in Rydberg atom systems. To overcome this limitation various approaches have been proposed, ranging from choosing anisotropic Rydberg states~\cite{glaetzle2014} to making use of multi-layer arrays~\cite{QSLtrimer}.
Here, we employ a method that avoids these experimental challenges by generalizing the concept of \textit{Rydberg gadgets} that have been developed to encode arbitrary classical constraints, e.g., for quantum optimization problems~\cite{Hannes_PRXQuantum.4.010316,PRXQuantum.3.030305,kim_rydberg_2022,Lanthaler_2023,HPB_PhysRevB.108.085138}. A gadget is a particular configuration of Rydberg atoms designed to provide a specific constraint solely as a result of the isotropic Rydberg blockade and the lattice geometry. 
While such constraints allow us to realize the frustration that can lead to the emergence of a QSL phase~\cite{HPB_PhysRevB.108.085138}, understanding if the accessible quantum fluctuations actually stabilize such a phase in models constructed from Rydberg gadgets is an open question. 

In this work, we propose simple gadgets that encode the constraint of dimer models and study the physics associated with the corresponding quantum fluctuations.
Quantum dimer models~\cite{moessner_quantum_2008,PhysRevLett.86.1881,PhysRevB.74.134301,PhysRevLett.61.2376,PhysRevB.64.144416,yan_triangular_2022} have recently sparked revived interest after the theoretical proposal~\cite{PhysRevX.11.031005} and corresponding experimental realization~\cite{doi:10.1126/science.abi8794} with a system of Rydberg atoms positioned on the edges of a kagome lattice, realizing what we call ``diluted dimer constraint'' (see below).
Here, we focus on geometries where this constraint is not natively provided by the Rydberg blockade (cf. Fig.~\ref{fig:illustration_encoding}(a-b)), namely the square and triangular lattices. We show that in the Rydberg gadget model, i.e., the system in which gadgets are placed on the vertices of such lattices, the diluted dimer constraint is recovered in the limit of large density of Rydberg excitations. We address the effect of quantum fluctuations via analytical perturbation theory and numerical exact diagonalization (ED) of the Rydberg atom Hamiltonian. We analyze the ground state phase diagrams of these systems and demonstrate the onset of $U(1)$ and $\mathbb{Z}_2$ QSLs, respectively on the square and triangular lattice. Moreover, by studying dynamical protocols to prepare them, we showcase the experimental relevance of our approach.

\paragraph{Diluted dimer models in Rydberg atom arrays. --} \label{sec:RydbergHamiltonian}

In a Rydberg atom array, neutral atoms are typically trapped into a configurable lattice geometry and approximated as two-level systems consisting of the ground state $|g\rangle$ and a highly-excited Rydberg state $|r\rangle$. Each atom is driven by a coherent laser field with Rabi frequency $\Omega$ and detuning $\Delta$. The Hamiltonian for this system is~\cite{RevModPhys.82.2313,browaeys_many-body_2020,PhysRevB.66.075128,PhysRevB.69.075106}  
\begin{equation}\label{eq:Rydberg}
    \hat{H}_{\text{Ryd}} =  \Omega \sum_{i}  \hat{\sigma}^{x}_{i}  - \sum_{i} \Delta_{i} \hat{n}_{i}  + \sum_{ij} U_{ij} \hat{n}_{i} \hat{n}_{j},
\end{equation}
where $\hat{\sigma}^x = |r\rangle\langle g|+|g\rangle\langle r|$, $\hat{n} = |r\rangle\langle r|$, and $i$ labels the position $\bm{x}_i$ of each atom. The interaction is of the van der Waals form $U_{ij} \sim |\bm{x}_{i}-\bm{x}_{j}|^{-6}$, which is strongly repulsive at short distances and rapidly decays at large distances. The strong distance dependence of the interaction strength allows for a blockade approximation, where interactions within the so-called blockade radius $R_\textrm{b}$ are assumed to be infinite and can be treated as the constraint $\hat{n}_{i} \hat{n}_{j}=0$ for $|\bm{x}_{i}-\bm{x}_{j}|<R_{\textrm{b}}$, while interactions at larger distances are neglected.

In the blockade approximation, for certain atom arrangements and a properly chosen $R_b$, the Rydberg atom system can be exactly mapped into a dimer model subject to the diluted dimer constraint, where \emph{at most} one dimer connects to each vertex of a lattice. The mapping consists of interpreting the atom positions as the midpoint of the edges of a lattice, and identifying a Rydberg excitation with the presence of a dimer on the corresponding edge. This correspondence however only holds on certain geometries, such as the ruby~\cite{PhysRevX.11.031005} and kagome~\cite{Ruben_sahay2023quantum} lattices, which are equivalent to dimer models on the kagome and honeycomb lattices, respectively. 

The effective Hamiltonian induced on the dimer degrees of freedom by Eq.~\eqref{eq:Rydberg} is given by $\hat{H} = \Omega \sum_{i}  \hat{\sigma}^{x}_{i}  - \Delta\sum_{i}  \hat{n}_{i}$, where $i$ labels the edges, and $\hat{\sigma}_i^x$ adds or removes dimers if allowed by the diluted dimer constraint. We dub this model diluted dimer model (DDM) and observe that it is well defined regardless of the mapping into a Rydberg atom system. In the limit $\Delta/\Omega\rightarrow\infty$, its low-energy subspace is spanned by the fully-packed dimer configurations with one dimer per vertex. At finite $\Delta/\Omega$, the off-diagonal term has two, competing effects on the DDM ground state: it favors coherence between maximal dimer coverings on the one hand, and lowers the density of dimers, on the other. Whereas the former typically stabilizes QSL phases, the latter could spoil it. Therefore, establishing the emergence of QSLs ground states in DDMs requires a non-perturbative treatment of quantum fluctuations, which we tackle below with ED methods.

In a generic lattice, the correspondence between Rydberg atom arrays in the blockade approximation and DDM does not hold. This is illustrated in Fig.~\ref{fig:illustration_encoding}(a) and (b) for the square and triangular lattices, respectively. For instance, on the square lattice (cf. Fig.~\ref{fig:illustration_encoding}(a)), the distance between atoms in the horizontal and vertical directions are both equal to the lattice spacing $a$. The diluted dimer constraint however would correspond to a blockade constraint that is active in one direction, but not in the other one, which is clearly incompatible with the isotropic blockade (see Fig.~\ref{fig:illustration_encoding}(a)).  
In what follows, we provide a solution to this issue by inserting gadgets on the lattice vertices (cf. Fig.~\ref{fig:illustration_encoding}(c-d)). These gadgets are designed to transmit the blockade of an excited atom on an edge to all the atoms on adjacent edges, but to no atom on any other edge.
We call the resulting model gadget model. Importantly, while this gadget model is designed to share similarities with the DDM on the same lattice, they differ in some key aspects. In particular, the diluted dimer constraint is enforced on the low-energy spectrum of the gadget model Hamiltonian only in the large $\Delta/\Omega$ regime.
Furthermore, the number of states in a unit cell is larger in the gadget model, resulting in an exponentially larger Hilbert space. Nevertheless, as we show below, the gadget model ground state satisfies the diluted dimer constraint in a wide parameter range and can even exhibit enhanced QSL features w.r.t. the DDM ground state. 
	
\paragraph{Diluted dimer models with Rydberg gadgets. --} \label{sec:Setup}

We now specify the Rydberg gadget models depicted in Fig.~\ref{fig:illustration_encoding}(c-f). We consider the square and triangular lattices, with a Rydberg atom on each edge, and a gadget on each vertex. The gadgets consist of 4 or 6 atoms on the square and triangular lattices, respectively. Edge and gadget atoms are represented by blue and orange dots, while the blockade radius is indicated by the green disk in Fig.~\ref{fig:illustration_encoding}(c,d). An edge atom blockades all the gadget atoms on adjacent vertices, except the two gadget atoms that are furthest away (cf. left side of  Fig.~\ref{fig:illustration_encoding}(c,d)). A gadget atom blockades all atoms in the same gadget and all edge atoms adjacent to that vertex, excluding the farthest one (cf. right side of Fig.~\ref{fig:illustration_encoding}(c,d)). This atom arrangement yields the dimer constraint as follows.
Whenever an atom in a gadget is excited to the Rydberg state, only atoms on the adjacent edges can be excited to the Rydberg state. This is akin to the constraint of having only one dimer on any edge connected to a vertex (cf. Fig.~\ref{fig:illustration_encoding}(e,f)). Accordingly, we call a gadget  \emph{active}, if it contains a Rydberg excitation. 
To favor active gadgets, the detuning for the gadget atoms, $\Delta_{\rm gad}$, is adjusted relative to the detuning of the edge atoms, $\Delta$, i.e. $\Delta_{\rm gad}=\alpha\Delta$. In the classical limit ($\Omega=0$) gadgets are active when $\Delta>0$ and $\alpha>3/2$ and $\alpha > 1$ for square and triangular lattices, respectively~\cite{SupMat}.
In summary, as shown in Fig.~\ref{fig:illustration_encoding}(e,f), a single dimer in the gadget model is represented by an excited atom on an edge, and it is always accompanied by two active gadgets at the neighboring vertices.
    
\begin{figure}
    \centering  
    \includegraphics[width=1.0\columnwidth]{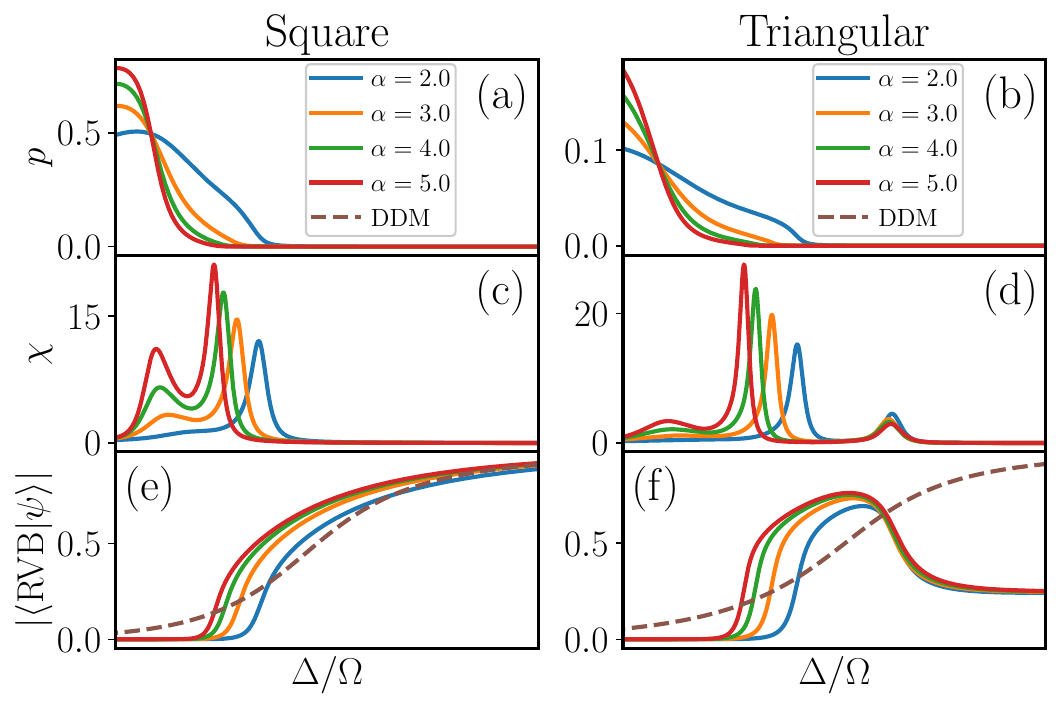}
    \caption{
        \textbf{Ground state phase diagram.} 
        Dimer-constraint violation probability (a,b), fidelity susceptibility Eq.~\eqref{eq:FS} (c,d), and ground state RVB overlap (e,f) as a function of $\Delta/\Omega$ and for different values of $\alpha = \Delta_\mathrm{gad}/\Delta$, on the square (a,c,e) and triangular lattices (b,d,f). The dashed lines indicate the diluted dimer model. Results are obtained on 10 and 8 unit-cell clusters for the square and triangular lattices, respectively~\cite{SupMat}.
        }
    \label{fig:GroundState}
\end{figure}

In this setup, the degenerate ground state subspace of the Rydberg Hamiltonian in the blockade approximation, and for $\Omega = 0$ and $\Delta>0$ is the same as in the associated DDM, and it is composed of the fully-packed dimer coverings. An important difference between Rydberg gadget models and DDMs is that the former allows for configurations in which there is more than one excited atom on edges connecting to the same vertex if the gadget on that vertex is inactive. However, as we will show below, when $\Delta \gtrsim 0$, the gadgets activate, thus penalizing these unwanted configurations at low energy (cf. Fig.~\ref{fig:GroundState}(a,b)). We will demonstrate that, although the gadget model is not guaranteed to reproduce the ground state properties of the DDM, it enables the observation of QSL features in the ground and dynamically prepared states. 

\paragraph{Ground state phase diagram. --} \label{sec:GroundState}

We performed ED calculations for the ground state of DDM and gadget models on finite systems with periodic boundary conditions. We employ a unit cell of 2 and 3 edges for the square and triangular lattices, respectively (cf. Fig~\ref{fig:illustration_encoding}(a,b)). Since there is one gadget per vertex, the gadget models for these two lattices have 6 and 9 atoms per unit cell, resulting in a much larger Hilbert space. Therefore, we focus on periodic clusters of up to 8 and 10 unit cells for square and triangular lattices~\cite{SupMat}. Our figure of merit for the characterization of a QSL is the overlap with the resonating valence bond (RVB) state~\cite{Giuliano_PhysRevLett.129.090401,QSLtrimer,quera2022trimer}, i.e., the equal-weight superposition of fully-packed dimer coverings $\{ |c\rangle \}$:
 \begin{equation}
     |\mathrm{RVB}\rangle = \frac{1}{\sqrt{\mathcal{N}}} \sum_{c} |c\rangle,
 \end{equation} 
 where $\mathcal{N}$ is the total number of these configurations. These states are the simplest representatives of QSLs and have very different properties on bipartite and non-bipartite lattices, like the square and triangular lattices. In the former case they encode the physics of a compact $U(1)$ deconfined gauge theory and have power-law decaying correlations, while in the latter they are described by a deconfined $\mathbb{Z}_2$ gauge theory and their correlations are short-range~\cite{moessner_quantum_2008,Savary_2017,doi:10.1142/S0217984990000295,PhysRevLett.89.137202,PhysRevB.65.024504,PhysRevB.69.224415,10.1093/acprof:oso/9780198785781.003.0003}.

To identify the boundaries of the gadget model phase diagram, we also consider the ground state fidelity susceptibility~\cite{PhysRevE.74.031123,doi:10.1142/S0217979210056335}
\begin{equation}
    \chi = -\frac{2}{\delta\lambda^2}\ln\left|\bra{\psi_0(\lambda)}\ket{\psi_0(\lambda + \delta\lambda)}\right|,
    \label{eq:FS}
\end{equation}
where $\lambda=\Delta/\Omega$, $|\psi_0(\lambda)\rangle$ is the ground state wavefunction, and $\delta\lambda$ is a small change in the parameter $\lambda$. Peaks in $\chi$ serve as a diagnostic tool for the location of phase transitions.

For the gadget model on the square lattices, $\chi$ exhibits two peaks, indicating two qualitative changes in the ground state (cf. Fig.~\ref{fig:GroundState}(c)). As we increase $\Delta/\Omega$, the system first undergoes a crossover from a trivially disordered regime to a ``gadget-active'' regime~\cite{Crossover}. As shown in Fig.~\ref{fig:GroundState}(a), this process is characterized by a sharp decrease in the ground state probability $p$ of violating the diluted dimer constraint, given e.g. by configurations with two dimers connecting to one vertex with an inactive gadget. Subsequently, the system undergoes a transition to a phase characterized by a significant overlap with the RVB state, as shown in Fig.~\ref{fig:GroundState}(e). Since $U(1)$ QSLs are expected to be unstable in 2D~\cite{polyakov1977}, we interpret this RVB phase as an emergence of a finite-size $U(1)$ QSL, which is not stable in the thermodynamic limit~\cite{Ruben_sahay2023quantum}. 

We note that the RVB overlaps for the DDM and gadget model on the square lattice converge to the same value in the classical limit $\Delta/\Omega\rightarrow\infty$, as can be seen from Fig.~\ref{fig:GroundState}(e). This result is readily understood through perturbation theory, which yields the same leading-order effective Hamiltonian for the two models~\cite{SupMat}
\begin{equation}
        \hat{H}_{\square}=-t\sum_r\Bigl(\KET[\Big]{\recy}\BRA[\Big]{\recx}+\textup{h.c.}\Bigr),
\end{equation}
where the thick lines correspond to dimers, and the flipping strength $t$ is proportional to $\Omega^{4}/\Delta^{3}$ and $\Omega^{12}/\Delta^{11}$ for the diluted and gadget model, respectively.
	
In Fig.~\ref{fig:GroundState}(d,f), we present the numerical results for the triangular lattice, where the fidelity susceptibility for the gadget model exhibits three peaks. Similarly to the square lattice, as $\Delta/\Omega$ increases, the system first undergoes a crossover from the trivially disordered regime to a gadget-active regime (cf. Fig.~\ref{fig:GroundState} (b)). On the triangular lattice, however, a high-RVB overlap phase is observed only at intermediate values of $\Delta/\Omega$, and a third peak in $\chi$ indicates a further transition to an ordered valence-bond solid (VBS) phase. We note that this is not the case for the DDM (dashed line in Fig.~\ref{fig:GroundState}(f)), where the RVB overlap monotonically increases with $\Delta/\Omega$. In contrast to the square lattice, $\mathbb{Z}_2$ QSLs can be stable on the triangular lattice. Therefore, this high-RVB overlap phase could indicate the emergence of a $\mathbb{Z}_2$ QSL phase that survives in the thermodynamic limit. The fact that RVB overlaps for diluted dimer and gadget models do not match in the limit $\Delta/\Omega \to \infty$ can again be understood via perturbation theory. The effective Hamiltonians for the two models are~\cite{SupMat}
\begin{equation}\label{eq:H_eff_tri}
    \begin{split}
        &\hat{H}_{\triangle,\mathrm{diluted}}=-t\sum_r\Bigl(\KET[\Big]{\plaqy}\BRA[\Big]{\plaqx}+\textup{h.c.}\Bigr),\\
        &\hat{H}_{\triangle,\mathrm{gad}}=-V\sum_r\Bigl(\KET[\Big]{\plaqy}\BRA[\Big]{\plaqy}+\KET[\Big]{\plaqx}\BRA[\Big]{\plaqx}\Bigr).
    \end{split}
\end{equation}
The leading-order term for the diluted model is the dimer flipping term from 4th order, with strength $t\propto\Omega^{4}/\Delta^{3}$. For the gadget model, instead, the leading order is a non-trivial diagonal term with strength $V\propto \Omega^6/\Delta^5$ arising at 6th order. In this case, the dimer flipping term only occurs at 12th order, with a strength proportional to $\Omega^{12}/\Delta^{11}$. This leads to the difference between the two RVB overlaps for $\Delta/\Omega \to \infty$.

We stress that the quantitative comparison of ground state RVB overlaps outlined in Fig.~\ref{fig:GroundState} (e) and (f) between DDMs and gadget models is remarkable. In fact, for fixed $\Delta/\Omega$, RVB overlaps in the gadget model can be larger than in the DDM despite the much larger Hilbert space dimension of the former~\cite{HS_size}. Moreover, the gadget system possesses the additional parameter $\alpha = \Delta_{\rm gad}/\Delta$, which can be tuned to increase RVB fidelities further and thus enhance QSL features.

\paragraph{Dynamical preparation. --} \label{sec:DynamicalPreparation}	

\begin{figure}
    \centering  
    \includegraphics[width=1.0\columnwidth]{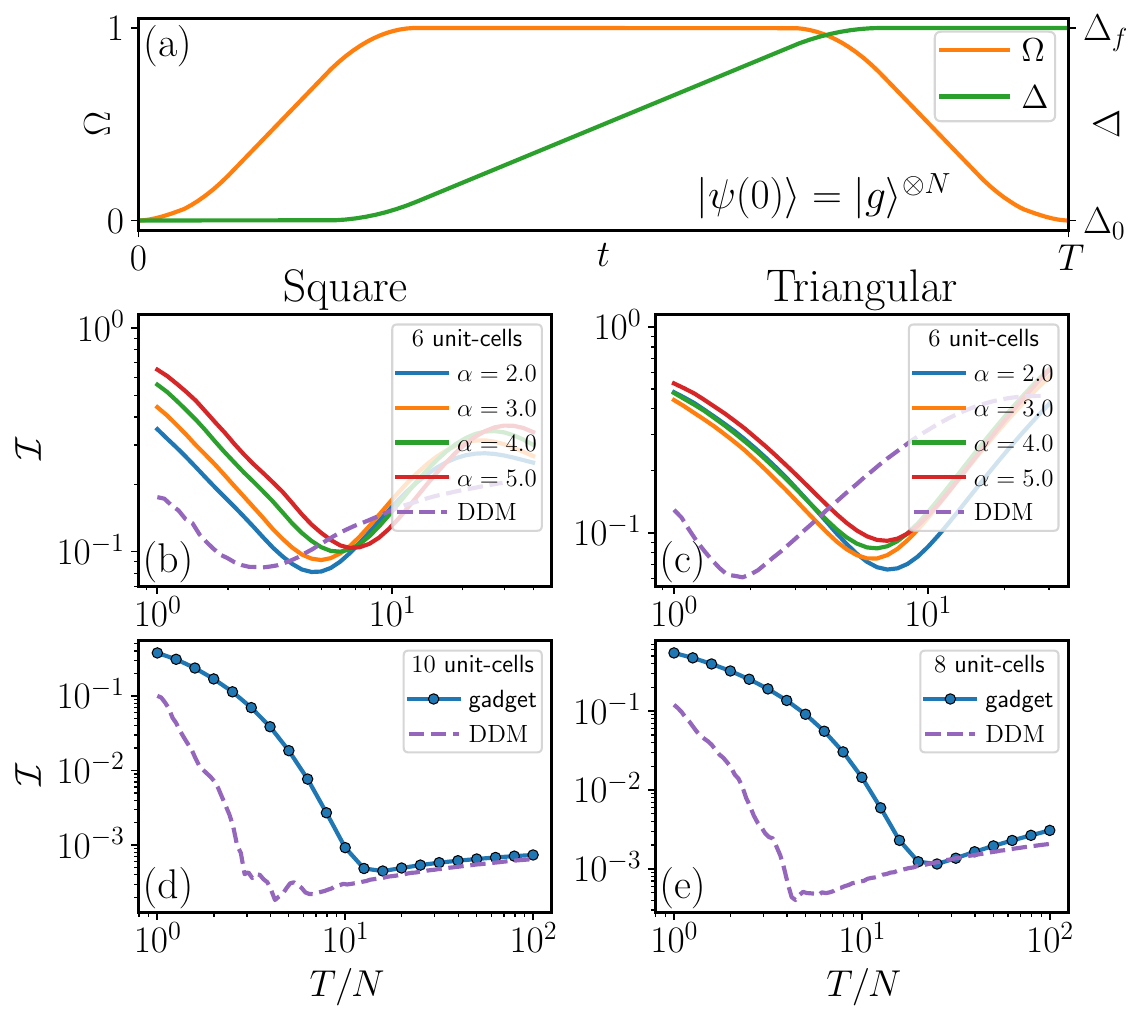}
    \caption{
        \textbf{RVB dynamical preparation infidelity.} (a) Dynamical preparation protocol. 
        (b,c) Infidelities $\mathcal{I} \equiv 1-|\langle\psi(T)|\mathrm{RVB}\rangle|$ obtained from periodic clusters with 6 unit cells as a function of $T/N$ with $T$ the total preparation time and $N$ the number of atoms. The dashed and solid lines are the infidelities obtained in the diluted dimer model and gadget model, for different values of $\alpha = \Delta_\mathrm{gad}/\Delta$ in the latter. 
        (d,e) Same as (b,c) but with $\alpha=2$, for clusters with 10 and 8 unit cells for the square and triangular lattices, respectively.} 
    \label{fig:DynamicalPreparation}
\end{figure}
    
We now analyze dynamical schemes typically discussed for preparing QSLs in experiments. 
We focus on the preparation protocol illustrated in Fig.~\ref{fig:DynamicalPreparation}(a). The atoms are initialized in their ground state, corresponding to the many-body ground state of the Hamiltonian Eq.~\eqref{eq:Rydberg} when $\Omega=0$ and $\Delta<0$. The Rabi frequency is then gradually increased to its maximum value, which we set to $1$ and fix our time units over the time interval $0 < t < T_{1}$. Following this, the detuning is linearly ramped up from an initial value $\Delta_0<0$ to a positive value $\Delta_{f}$ within the time span $T_{1} < t < T_{1} + T_{2}$. Finally, the Rabi frequency is gradually decreased to zero. Here, we set $T_2 = 0.5T$, $T_1 = T_3 = 0.25 T$, $\Delta_0 = -5$, and $\Delta_f = 1.5$~\cite{Giuliano_PhysRevLett.129.090401}. 

When the annealing time $T$ goes to infinity, the system evolves adiabatically and remains in the instantaneous ground state. The final state ends up in the ground state obtained in the classical limit $\Delta/\Omega\rightarrow\infty$, implying that the RVB overlap of the final state for $T \to \infty$ has to approach the RVB overlap of the ground state for $\Delta/\Omega \to \infty$ (cf. Fig.~\ref{fig:GroundState} (e,f)). Conversely, for small $T$, only a few atoms are excited throughout the dynamics, resulting in a significant number of defects and a suppressed RVB overlap. Our primary interest is on the semi-adiabatic preparation in the intermediate regime of total preparation times $T$, where RVB fidelities are expected to be maximal~\cite{Giuliano_PhysRevLett.129.090401}. In Fig.~\ref{fig:DynamicalPreparation}(b-e), we plot the infidelity between the final state and the RVB state, i.e., $\mathcal{I} \equiv 1-|\langle\psi(T)|\mathrm{RVB}\rangle|$, for the square and triangular lattices. In both cases, we not only observe an enhanced RVB overlap at intermediate $T$ w.r.t. the adiabatic case, but also an optimal overlap which is comparable in the DDMs and gadget models, reaching $\mathcal{I} < 10^{-3}$ for the largest systems considered. In Fig.~\ref{fig:DynamicalPreparation}(b)-(c), we demonstrate how the extra control parameter $\alpha$ of the gadget model can be used to further minimize $\mathcal{I}$ \cite{SupMat}. Due to the large Hilbert space dimension, we did not perform such optimization in the larger periodic clusters in Fig.~\ref{fig:DynamicalPreparation} (d,e).

\paragraph{Conclusions. --} \label{sec:conclusion} 
We proposed a method based on Rydberg gadgets to engineer the constraint of dimer models in neutral atom arrays in the regime of maximal density of Rydberg excitations. This method can be realized in experiments simply by placing the gadgets on the vertices of a lattice. We analyzed the effect of quantum fluctuations on the gadget model by numerically studying the ground state phase diagram on the square and triangular lattices, and compared it to the simpler diluted dimer models, which are not naturally realized with an isotropic Rydberg blockade in these geometries.
Making use of perturbation theory, we highlighted similarities and differences between DDMs and gadget models, and showed that the ground state of the latter can exhibit enhanced QSL features w.r.t. their diluted dimer counterparts. 
Finally, we demonstrated that through non-adiabatic dynamical protocols, it is possible to achieve high-fidelity preparation of $U(1)$ and $\mathbb{Z}_2$ spin liquid states on the square and triangular lattice, respectively.

Our witness for topological order --the RVB overlap-- is purely theoretical and cannot be directly probed in experiments. While non-local string operators have been devised to detect such QSLs in experiments~\cite{PhysRevX.12.041029,doi:10.1126/science.abi8794}, generalizing these operators in our gadget models can be challenging. However, alternative observables, such as the dimer density-density correlation function and the dimer structure factor~\cite{Ruben_sahay2023quantum,yan_triangular_2022}, can help distinguish topological RVB states from trivial ones. Another potential approach involves coherently transferring the population from the Rydberg state to a non-interacting hyperfine state\cite{Bluvstein_2022}, enabling state tomography on small systems.

In this work, we considered the blockade approximation of Rydberg Hamiltonians and ignored the long-range interaction tails, which is a crude approximation for real experiments. However, this approximation can be improved by properly tuning the blockade radius and utilizing different atomic species~\cite{SupMat}. Although we only applied Rydberg gadgets to obtain the dimer constraint on square and triangular lattices, this method can be generalized to a larger class of models and used to engineer other kinds of constraints which, when combined with quantum fluctuations, could provide more exotic forms of topological order~\cite{HPB_PhysRevB.108.085138}. Specifically, we discuss the Rydberg gadgets for $\mathcal{Z}_2$ lattice gauge theory with dynamical matter in Supplementary Material~\cite{SupMat}.

\let\oldaddcontentsline\addcontentsline
\renewcommand{\addcontentsline}[3]{}
\begin{acknowledgments}
\paragraph{Acknowledgements. --}
We are grateful to Giacomo Giudice for providing support on part of the code used for the numerical simulations, and to Robert Ott for providing feedback on the manuscript.
We thank Tao Shi, Hongzheng Zhao, Robert Ott, and Torsten Zache for valuable discussions.
G.G. acknowledges support from the European Union’s Horizon Europe program under the Marie Sklodowska Curie Action TOPORYD (Grant No. 101106005).
This work is supported by the European Union’s Horizon Europe research and innovation program under Grant Agreement No. 101113690 (PASQuanS2.1), the ERC Starting grant QARA (Grant No.~101041435), the EU-QUANTERA project TNiSQ (N-6001), and by the Austrian Science Fund (FWF)[grant DOI 10.55776/COE1]. 
\end{acknowledgments}

\bibliography{library}	

\begin{thebibliography}{74}
\bibitem [{Crossover()}]{Crossover}%
  \BibitemOpen
  \bibinfo {note} {We expect this ground state change to remain smooth in the thermodynamic limit. We validated this scenario by noting that the peak in the fidelity susceptibility per atom does not scale with increasing system size, as expected from a true thermodynamic limit phase transition.}\BibitemShut {Stop}%
\bibitem [{HS_size()}]{HS_size}%
  \BibitemOpen
  \bibinfo {note} {For the clusters considered in Fig.~\ref{fig:GroundState}, the dimensions of the constrained Hilbert space for the diluted dimer model are approximately $2^{9.6}$ and $2^{9.0}$ for square and triangular lattices, respectively. For the gadget models, the dimensions are approximately $2^{28.2}$ and $2^{25.2}$ for square and triangular lattices, respectively.}\BibitemShut {Stop}%
\end{thebibliography}
\let\addcontentsline\oldaddcontentsline

\clearpage
\onecolumngrid
\begin{center}
    \textbf{\Large Supplementary Material}
\end{center}
\normalsize

\setcounter{equation}{0}
\setcounter{figure}{0}
\setcounter{table}{0}
\makeatletter
\renewcommand{\theequation}{S\arabic{equation}}
\renewcommand{\thefigure}{S\arabic{figure}}
\setlength\tabcolsep{10pt}
\setcounter{secnumdepth}{2}

\newcommand\numberthis{\addtocounter{equation}{1}\tag{\theequation}}
\newcommand{\insertimage}[1]{\includegraphics[valign=c,width=0.04\columnwidth]{#1}}

\tableofcontents

\section{Atom arrangement and clusters}\label{app:cluster}

\begin{figure}[h!]
    \centering  
    \includegraphics[width=0.8\columnwidth]{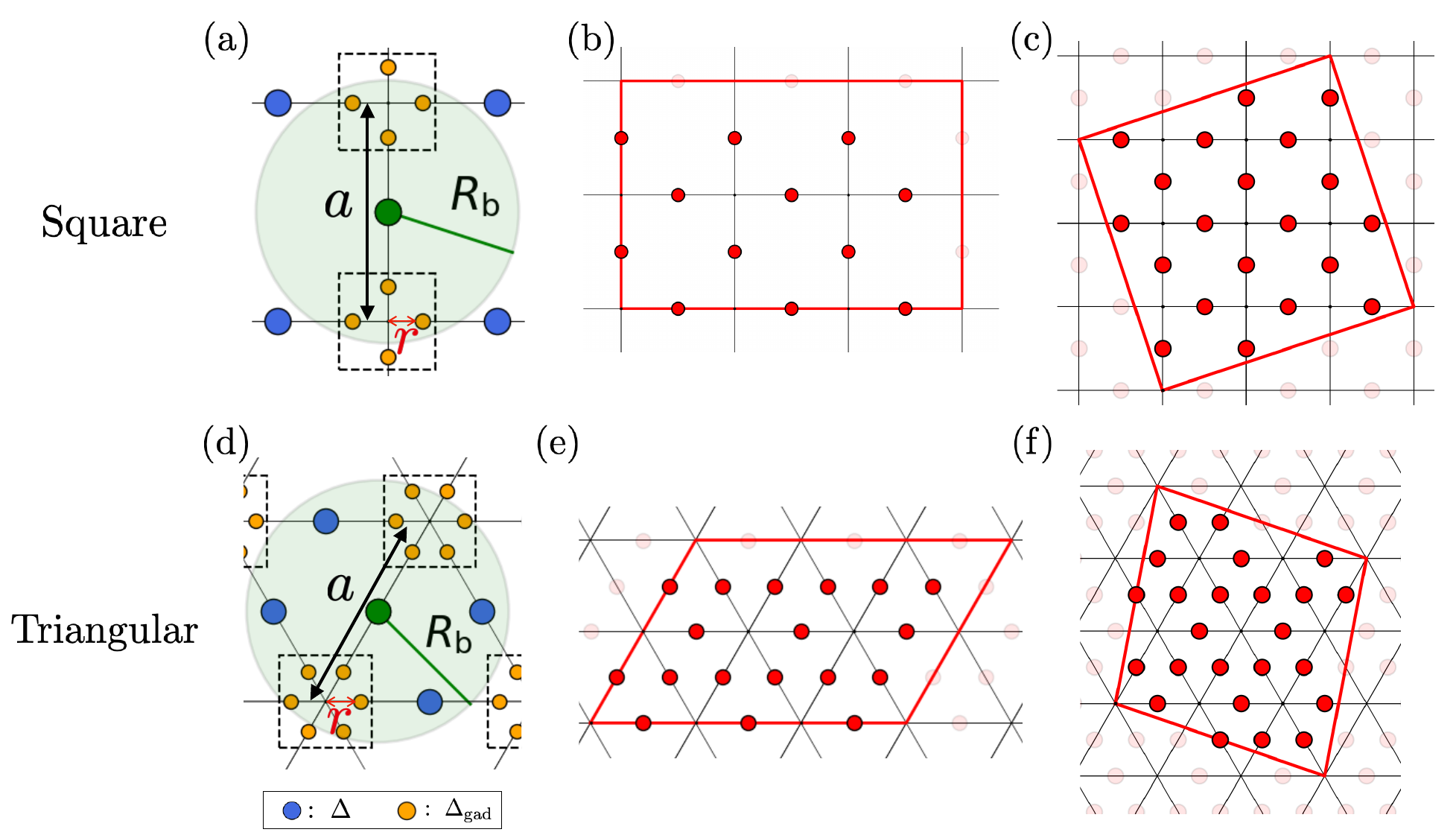}
    \caption{
        \textbf{Atom arrangement and clusters for ED calculations.} (a,d) Depiction of the atom arrangement for the square (upper) and triangular (bottom) lattices. (b,e) Clusters with $6$ unit-cells for the square (upper) and triangular (bottom) lattices. (c) Square lattice cluster with $10$ unit-cells. (f) Triangular lattice cluster with $8$ unit-cells.} 
    \label{fig:ED_cluster}
\end{figure}

 In this section, we give more details about the geometric setup used in this work. In the gadget model, the distance between the gadget atoms and their corresponding vertices is denoted as $r$. The lattice spacing is denoted as $a$ (cf. Fig.~\ref{fig:ED_cluster}(a,d)). 
 On the square lattice, the gadget method requires: ({\it i}) the edge atom {\it not} to be blockaded with the farthest gadget atoms in the two corresponding vertices ($R_{\mathrm{b}}<r+a/2$); ({\it ii}) the edge atom to blockade the other gadget atoms in the two corresponding vertices ($\sqrt{r^2+(a/2)^2}<R_{\mathrm{b}}$); and ({\it iii}) the atoms in different gadgets {\it not} to blockade each other ($R_{\mathrm{b}}<a-2r$). These conditions imply 
 \begin{equation}
R_{\min} = \sqrt{r^{2}+\left(\frac{a}{2}\right)^2} < R_b < \min \left( a-2r,r+\frac{a}{2}\right) = R_{\max}.
\label{eq:rb_square}
 \end{equation}
 Since our numerical simulations are in the blockade approximation with no long-range interaction tails, the choice of $r/a$ and $R_b$ is irrelevant, as long as the exact blockade is imposed on atoms at a distance shorter than $R_b$, with $R_b$ satisfying Eq.~\eqref{eq:rb_square}. However, for this to be a good approximation of the full Rydberg Hamiltonian we want to maximize the ratio $R_{\max}/R_{\min}$, such that $U(R_{\min})/U(R_{\max}) \gg 1 $. For this ratio to be maximal one must set $r=a/6$, which yields $U(R_{\min} )/U(R_{\max} ) = \left(4/\sqrt{10}\right)^6 \simeq 4.1$. 
 
 On the triangular lattice similar conditions hold: ({\it i}) $R_{\mathrm{b}}<r+a/2$; ({\it ii}) $\sqrt{r^2+(a+\sqrt{3}r)^2}/2<R_{\mathrm{b}}$; and ({\it iii}) $R_{\mathrm{b}}<a-2r$. Therefore, $R_b$ must satisfy
 \begin{equation}
 R_{\min} = \frac{1}{2} \sqrt{  r ^{2} + \left( a + \sqrt{3} r \right)^2 } < R_b < \min \left( a - 2r, r+\frac{a}{2} \right) = R_{\max}.
 \end{equation}
As in the square lattice case, the ratio $R_{\max}/R_{\min}$ is maximal when $r/a = 1/6$, which yields $U(R_{\min} )/U(R_{\max} ) = \left(4/\sqrt{10 + 3 \sqrt{3}}\right)^6 \simeq 1.2$. Differently from the square lattice, the strongest interaction above the blockade radius is not much smaller than the weakest interaction below the blockade radius, making the blockade approximation less justified. However, as we show in Sec.~\ref{app:exp_imp}, this issue can be mitigated by making use of multi-species arrays. 

 The exact diagonalization calculations are performed on finite-size systems with periodic boundary conditions.  Employing the blockade approximation, we truncate the full Hilbert space by excluding states that violate the blockade constraint. This reduction to the blockade-constrained Hilbert space enables us to simulate relatively large systems.
 The clusters we consider are shown in Fig.~\ref{fig:ED_cluster}. Note that the clusters are the same for both the gadget and non-gadget models. In the gadget models, each unit cell contains 6 atoms on the square lattice and 9 atoms on the triangular lattice.

\section{Relation between detuning on the edges and on the gadgets}\label{app:alpha_requirement}

For realizing the dimer model constraint in the gadget model the detuning on the edges $\Delta$ and on the gadgets $\Delta_\mathrm{gad}$ have to satisfy the relation $\Delta_\mathrm{gad} = \alpha \Delta$ with $\alpha>3/2$ and $\alpha > 1$ for the square and triangular lattices, respectively. These minimum values of $\alpha$ ensure the gadgets' activation when $\Delta > 0$. They can be inferred by comparing the energy density $\varepsilon_0$ of the dimer fully-packed configurations (i.e., one dimer connects to each vertex) with the energy density $\varepsilon_1$ of the other ground state candidates configurations, when $\Omega = 0 $.
For both the square and triangular lattices we have $\varepsilon_0 = - (1/2+\alpha)\Delta$.

\begin{figure}[h]
    \centering  
    \includegraphics[width=0.5\columnwidth]{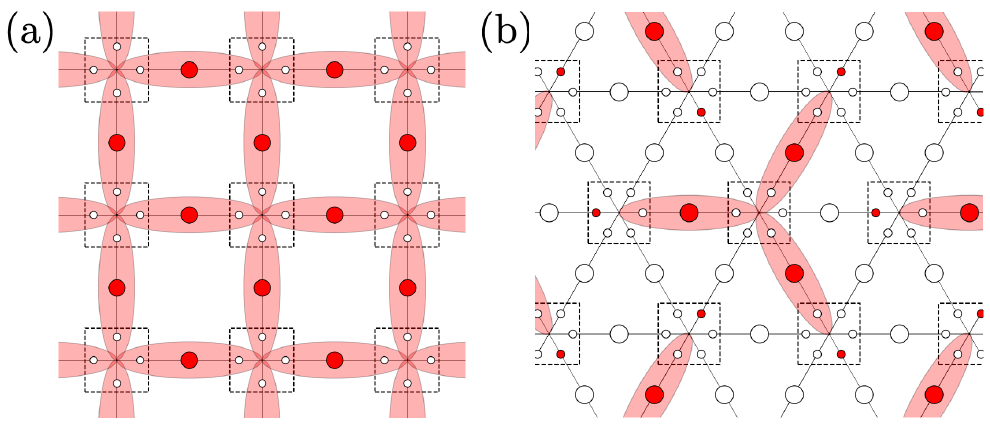}
    \caption{
        \textbf{Low-energy configurations in the gadget models.} (a) Gadget fully-inactive configuration on the square lattice. (b) ``Fan'' configurations on the triangular lattice.} 
    \label{fig:App_alpha}
\end{figure}

On the square lattice, as $\alpha$ is varied, the dimer fully-packed subspace competes with the unique gadget fully-inactive configuration shown in Fig.~\ref{fig:App_alpha}(a), where all edge atoms are in the Rydberg state and all gadget atoms are in the ground state. The energy density of such configurations is $\varepsilon_1 = - 2\Delta$. The condition $\varepsilon_1 > \varepsilon_0$ thus implies $\alpha > 3/2$.

On the triangular lattice, due to the direct blockade between nearest-neighbor edge atoms, the states competing with the dimer-fully packed subspace consist of the ``fan'' configurations depicted in Fig.~\ref{fig:App_alpha}(b). Their energy is $\varepsilon_1 = - 3(1+\alpha)\Delta/4$, so that one must set $\alpha > 1$ to have $\varepsilon_1 > \varepsilon_0$.

\section{Derivation of the effective Hamiltonian}\label{app:perturbation}
    
\begin{figure}[h]
    \centering  
    \includegraphics[width=0.8\columnwidth]{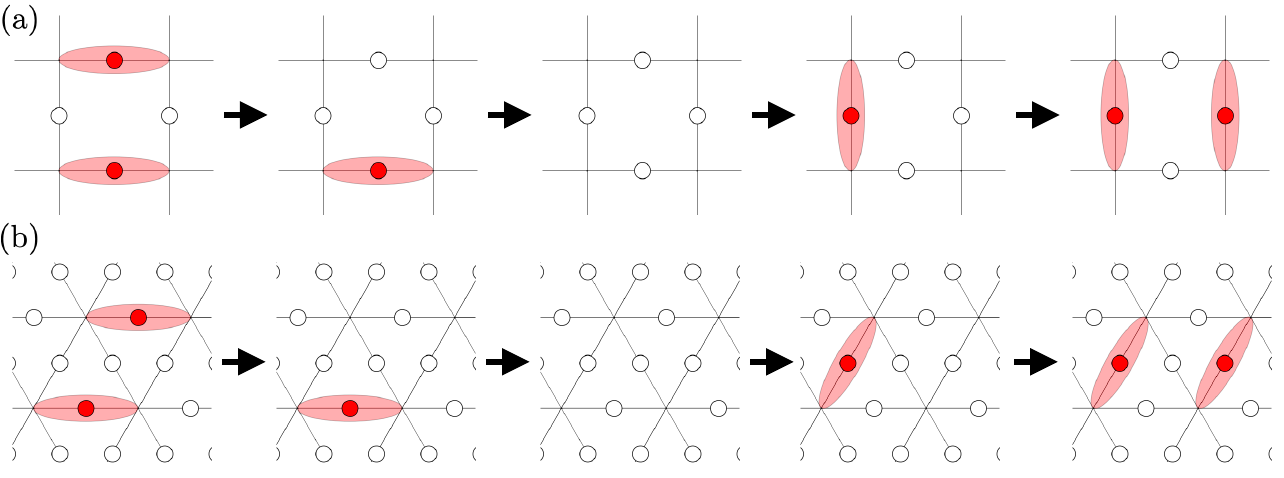}
    \caption{
        \textbf{Dimer flipping process in the diluted dimer model.} (a) Square lattice. (b) Triangular lattice. These 4th-order processes in perturbation theory contribute to a dimer-flipping term with strength $t\propto\Omega^4/\Delta^3$ to their effective Hamiltonians. Similar processes also exist in the gadget model, but they require to remove and add back the gadget atoms, and thus arise at 12th order, with strength $t\propto \Omega^{12}/\Delta^{11}$. } 
    \label{fig:App_Pert_dilute}
\end{figure}

In the blockade approximation, where the blockade constraint is implicitly assumed, the Hamiltonian reduces to an on-site form, which can be split into a diagonal term $\hat{H}_0=-\sum_i \Delta_i \hat{n}_i$, where $\Delta_i = \Delta$ for edge atoms and $\Delta_i = \alpha\Delta$ for gadget atoms, and an off-diagonal term $\hat{V}=\Omega \sum_i \hat{\sigma}_i^x$. We take the former as the unperturbed Hamiltonian $\hat{H}_0$. Its ground space is degenerate with energy $E_0$, and is made of all the fully-packed dimer configurations. The off-diagonal term contributes to the creation or destruction of a dimer.
The effective Hamiltonian at leading order $n$ is written as~\cite{Mila_2010}:
\begin{equation}
    \hat{H}_{\mathrm{eff}} \sim \hat{P}\hat{V} \left( \hat{Q}\frac{1}{E_0 - \hat{Q}\hat{H_0}\hat{Q}}\hat{Q}\hat{V} \right)^{n-1} \hat{P},
\end{equation}
where $\hat{P}$ is the projector onto the ground space of $\hat{H}_0$ (the dimer fully-packed states), and $\hat{Q}=1-\hat{P}$.

For diluted dimer models on both square and triangular lattices, the lowest-order non-trivial contribution is the flipping of a pair of dimers arising at order $n=4$, as shown in Fig.~\ref{fig:App_Pert_dilute}. The flipping strengths are proportional to $\Omega^4/\Delta^3$. In the gadget model, the analogous flipping term originates at order $n=12$, with strength $\Omega^{12}/\Delta^{11}$. In this case, the higher order of $\hat{V}$ is needed to enable the dimer to move by changing the Rydberg state position on the 4 gadgets involved. Since each gadget requires two flips of the atomic state, there are in total 8 extra applications of the off-diagonal perturbation w.r.t. the diluted dimer model. Note that a single-dimer flipping term could exist in finite-size systems with one narrow direction, such as the clusters in Fig.~\ref{fig:ED_cluster}(b,e).

\begin{figure}[h]
    \centering  
    \includegraphics[width=1.0\columnwidth]{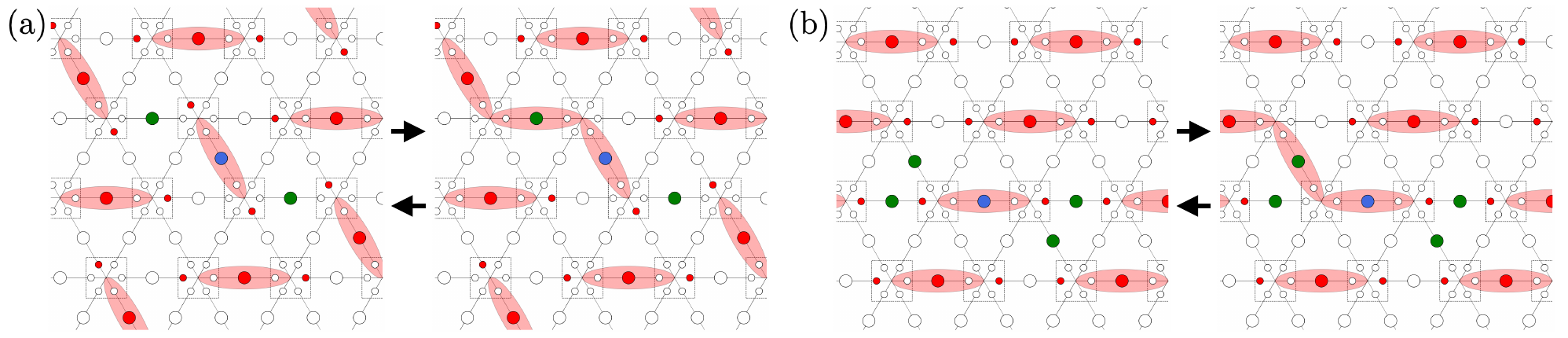}
    \caption{
        \textbf{Example of non-trivial diagonal processes in the triangular gadget model.} In these processes, one edge dimer can be created after removing two nearby gadget atoms. We focus on the dimer marked with blue dots. The allowed nearby edges are marked with green dots. 
        For one dimer in the staggered configuration (a)(left), two such processes (one of which is shown in (a)) contribute to the diagonal matrix element. In the columnar configuration (b)(left), there are four such processes, one of which is shown in (b).} 
    \label{fig:App_Pert_dia_6th}
\end{figure}

In the triangular lattice gadget model, the non-trivial diagonal term in Eq.~(6) of the main text results from the direct blockade between atoms on the edges. In fact, the edge atoms blockade not only the gadget atoms but also the 4 atoms on the nearby edges (cf. Fig.~1(d)). This direct blockade between edge atoms forbids certain intermediate processes in perturbation theory. For instance, for the staggered configuration shown in Fig.~\ref{fig:App_Pert_dia_6th}(a), two edge atoms can be excited after de-exciting two gadget atoms. This process yields back the initial configuration and thus constitutes a 6th-order diagonal process. Another example is the columnar configuration, for which more processes are allowed, as shown in Fig.~\ref{fig:App_Pert_dia_6th}(b). 
In the more general case, a single dimer (an excited edge atom) has 10 nearby edges, and 4 of them are directly blockaded by the edge atom (cf. Fig.~1(d)). We refer to these 4 edges as forbidden edges. When two dimers form a pair on a plaquette (e.g. $\plaqx$), they share one forbidden edge. In this case, a dimer can be created on one more nearby edge w.r.t. local configurations in which the dimers do not form a pair. Consequently, the pair configurations allow for one more diagonal process at 6th order in perturbation theory, and thus have lower energy compared to configurations where the plaquette is not flippable. This contributes a non-trivial diagonal term that counts the number of flippable plaquettes, with strength $V\propto\Omega^{6}/\Delta^{5}$.

\section{Rydberg excitation density}\label{app:density}

 \begin{figure}[h]
    \centering  
    \includegraphics[width=0.55\columnwidth]{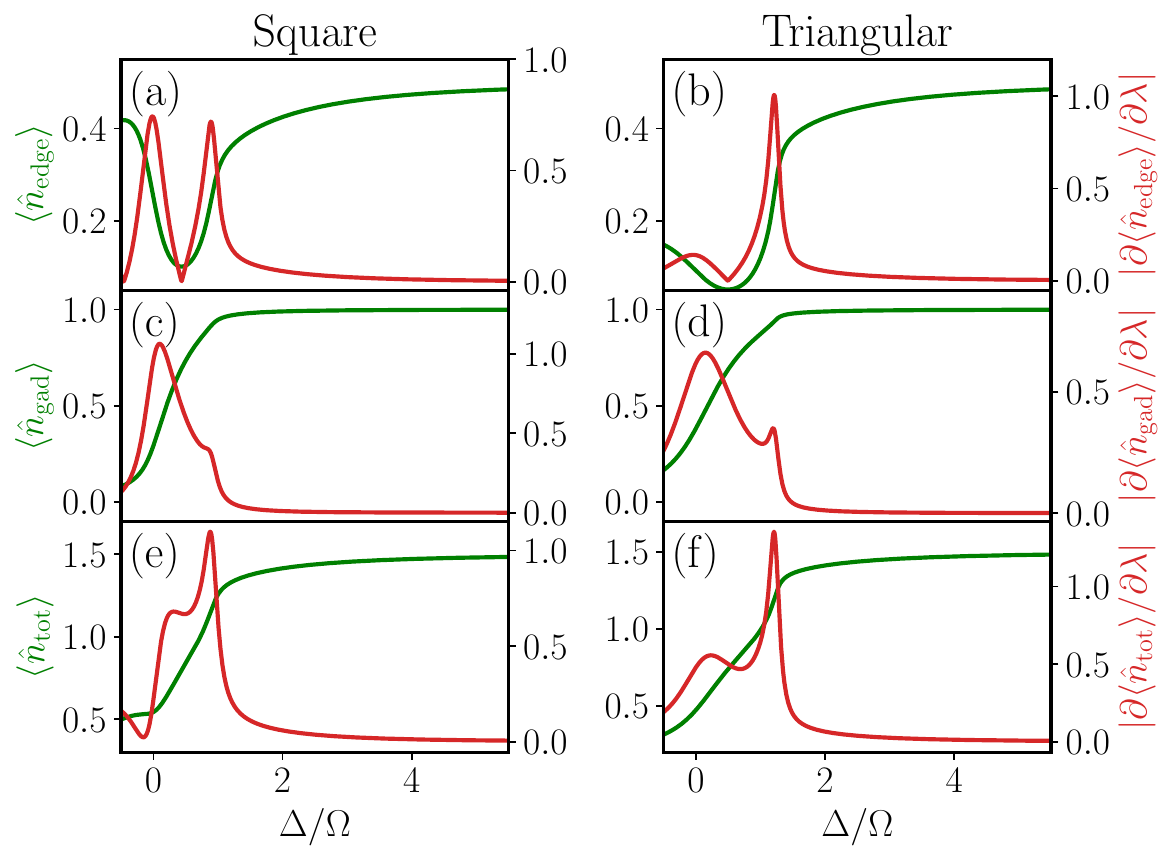}
    \caption{
        \textbf{Ground state Rydberg excitation density.} 
        Rydberg excitation densities (green lines) and their derivatives (red lines) for atoms on the edges (a,b), on the gadgets (c,d), and in total (e,f), in the ground state of the gadget model. We set $\alpha=\Delta_{\rm gad}/\Delta=5$. 
    }
    \label{fig:App_GroundState_N}
\end{figure}

The gadget and edge densities, namely the density of Rydberg excitations on the edges, can serve as another observable to witness the ``gadget-active'' regime, which we characterized in Fig.~2 of the main text with a sharp suppression of the ground state amplitude on atom configurations that violate the diluted dimer constraint. In Fig.~\ref{fig:App_GroundState_N}, we plot edge, gadget, and total densities in both gadget models as green solid lines. 
In the crossover from the disordered phase to the ``gadget-active'' regime, the gadget density quickly increases, while the edge density is suppressed due to the blockade between edge and gadget atoms. As $\Delta/\Omega$ further increases, other transitions arise, while the gadgets remain active. Eventually, both edge and gadget densities saturate to their maximal value. We note that, although the gadget density is not maximal for intermediate $\Delta/\Omega$, the gadgets accomplish their function of penalizing configurations that do not satisfy the diluted dimer constraint. This fact indicates that Rydberg gadgets in this setup are stable against gadget density fluctuations. 
In Fig.~\ref{fig:App_GroundState_N}, we also plot the densities derivatives of $\lambda=\Delta/\Omega$ as red solid lines, and show that they correctly signal the various phase transitions discussed in the main text.

\section{Optimization of infidelity with respect to the detuning ratio $\alpha$}\label{app:alpha}

\begin{figure}[h]
    \centering  
    \includegraphics[width=0.55\columnwidth]{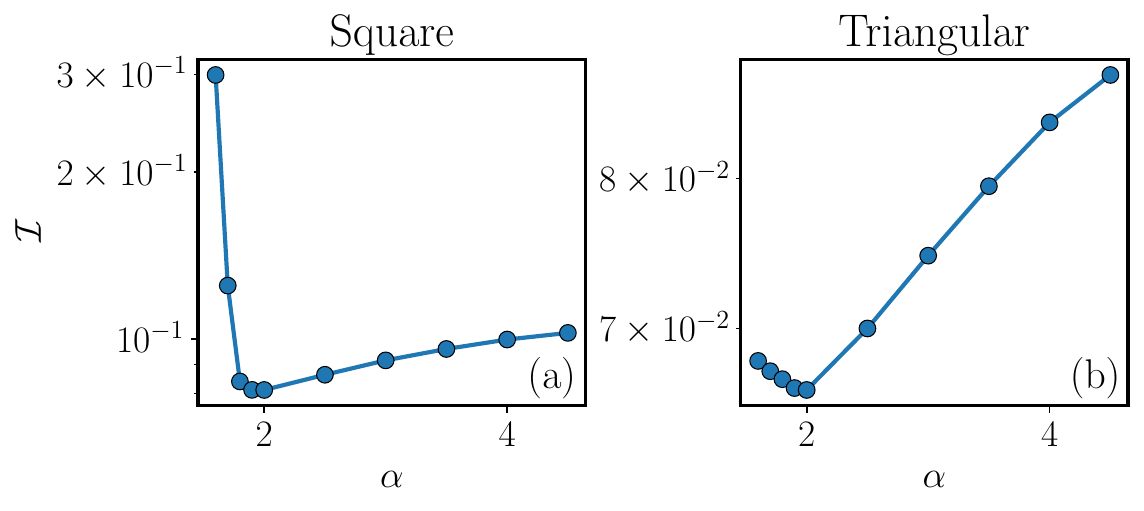}
    \caption{
        \textbf{Infidelities as a function of the detuning ratio $\alpha$.} Infidelities $\mathcal{I}$ obtained from periodic clusters with $6$ unit cells for (a) square and (b) triangular lattices. All other parameters are identical to those used in Fig.~3.} 
    \label{fig:App_optimize_alpha}
\end{figure}

We demonstrate that the control parameter $\alpha=\Delta_\mathrm{gad}/\Delta$ of the gadget model can be used to further minimize the infidelity $\mathcal{I}$ in dynamical preparation (cf.~Fig.~3). 
The detuning ratio $\alpha$ has a dual effect: on the one hand, a larger $\alpha$ enhances the constraint, ensuring the lowest energy subspace corresponds to the fully-packed configurations. On the other hand, increasing $\alpha$ reduces the strength of the dimer flipping term (cf Eq.~(5)), which weakens the coupling between different dimer configurations. Conversely, a smaller $\alpha$ can increase the dimer flipping strength but relax the constraint. Consequently, there exists an optimal $\alpha$ that minimizes the infidelity $\mathcal{I}$.

In Fig.~\ref{fig:App_optimize_alpha}, we plot the infidelity as a function of $\alpha$. As anticipated, the optimal $\alpha$ lies in an intermediate range, around $\alpha\approx2$. As discussed in Sec.~\ref{app:cluster}, $\alpha>3/2$ and $\alpha > 1$ are required for square and triangular lattices, respectively (cf. Sec.~\ref{app:alpha_requirement}).

\section{Improving the blockade approximation by multi-species arrays}\label{app:exp_imp}

\begin{figure}[h]
    \centering  
    \includegraphics[width=0.55\columnwidth]{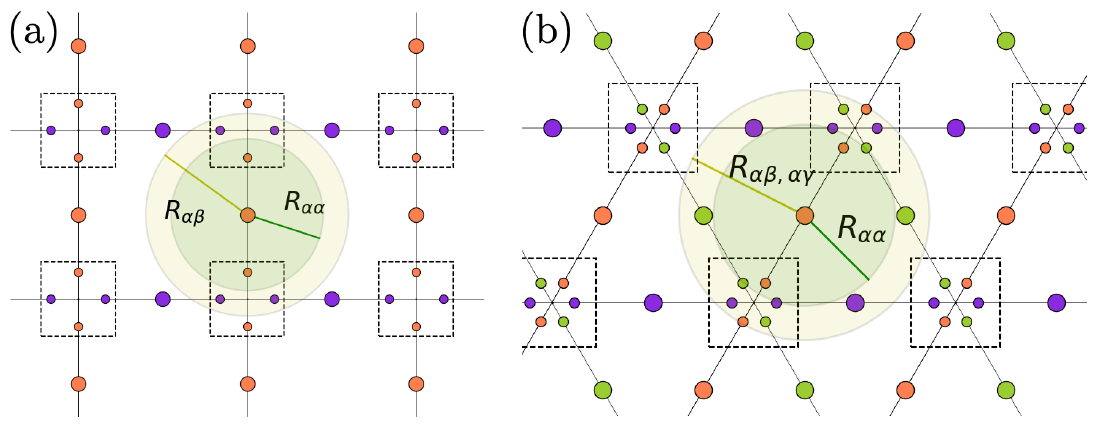}
    \caption{
        \textbf{Improvement by using multi-species atoms.} The different atom species are marked by different colors. (a) Square lattice. We set $r=0.16a$, $R_{\alpha\alpha}=0.45a$, and $R_{\alpha\beta}=0.6a$. (b) Triangular lattice. We set $r=0.11a$, $R_{\alpha\alpha}=0.45a$, and $R_{\alpha\beta}=0.62a$.} 
    \label{fig:App_Improve1}
\end{figure}	

In the main text, we assumed the blockade approximation of the van der Waals interaction, which implies $U(R_{\min} )/U(R_{\max} ) \gg 1$, where $U(R_{\min} )$ ($U(R_{\max} )$) are the strongest (weakest) interaction above (below) the blockade radius. As we discussed in Sec.~\ref{app:cluster}, this approximation is justified for the gadget model on the square lattice, where $U(R_{\min} )/U(R_{\max} ) \simeq 4.1$, but not on the triangular lattice,  where $U(R_{\min} )/U(R_{\max} ) \simeq 1.2$. In this section, we discuss how to mitigate this issue by employing multi-species atom arrays~\cite{bernien_twospecies,Singh_2023,anand2024dualspecies}. A similar effect can be obtained via multi-layer arrays~\cite{barredo_synthetic_2018,kim_rydberg_2022,PhysRevLett.130.180601}.
 
On the square lattice, we make use of two species of atoms, denoted as $\alpha$ and $\beta$. The blockade radii are denoted as $R_b^{\alpha\alpha}$, $R_b^{\alpha\beta}$, and $R_b^{\beta\beta}$, and we assume $R_b^{\alpha\alpha} = R_b^{\beta\beta}$. The atom arrangement and blockade radii are illustrated in Fig.~\ref{fig:App_Improve1}(a), where $\alpha$ and $\beta$ atoms are represented by orange and purple dots, respectively. The gadget requirements in terms of the blockade radii are 
\begin{align}
   & R_{\min}^{\alpha \alpha} =  \max \left(2r,\frac{a}{2}-r \right) < R_b^{\alpha\alpha}< \frac{a}{2} +r = R_{\max}^{\alpha \alpha}  , \\[1mm]
   & R_{\min}^{\alpha \beta}  = \sqrt{r^2+ \left( \frac{a}{2} \right)^2} < R_b^{\alpha\beta} < \sqrt{\left(a-r\right)^2+\left( \frac{a}{2} \right)^2} = R_{\max}^{\alpha \beta} .
\end{align} 

For the blockade approximation to be valid we want $r_{\alpha \alpha} = U(R_\mathrm{max}^{\alpha \alpha})/U(R_\mathrm{min}^{\alpha \alpha}) \gg 1$ and $r_{\alpha \beta} =  U(R_\mathrm{max}^{\alpha \beta})/U(R_\mathrm{min}^{\alpha \beta}) \gg 1$.
We can try to maximize both $r_{\alpha\alpha}$ and $r_{\alpha\beta}$ by choosing $r/a=0.16$. We get $r_{\alpha\alpha}\simeq 53$ and $r_{\alpha\beta}\simeq 41$, which are much larger than in the single-species setup ($r \simeq 4$, cf. Sec.~\ref{app:cluster}). 

A similar strategy applies to the triangular lattice. However, in this case, we need to introduce three species of atoms, $\alpha$, $\beta$, and $\gamma$. 
Assuming $R_{\alpha\alpha} = R_{\beta\beta} = R_{\gamma\gamma}$ and $R_{\alpha\beta} = R_{\alpha\gamma} = R_{\beta\gamma}$, the gadget requirements are

\begin{align}
   & R_{\min}^{\alpha \alpha} =  \max \left( 2r,\frac{a}{2}-r \right) < R_b^{\alpha\alpha}< \frac{a}{2} +r = R_{\max}^{\alpha \alpha}  , \\[1mm]
   & R_{\min}^{\alpha \beta}  = \frac{1}{2} \sqrt{r^2+\left(a+\sqrt{3}r\right)^2} < R_b^{\alpha\beta} < \frac{1}{2} \sqrt{r^2+3\left(a-r\right)^2} = R_{\max}^{\alpha \beta} .
\end{align} 

Again, we want to maximize both the intra-species interaction ratio $r_{\alpha\alpha}$ and the inter-species ratio $r_{\alpha\beta}$. If we set $r/a=0.11$, we have $r_{\alpha\alpha}\simeq 6.9$ and $r_{\alpha\beta}\simeq 7.5$, which are a notable improvement w.r.t. the single-specie setup ($r \simeq 1.2$, cf. Sec.~\ref{app:cluster}).

\section{Effect of the van der Waals interaction tail}

\begin{figure}[h]
    \centering  
    \includegraphics[width=0.52\columnwidth]{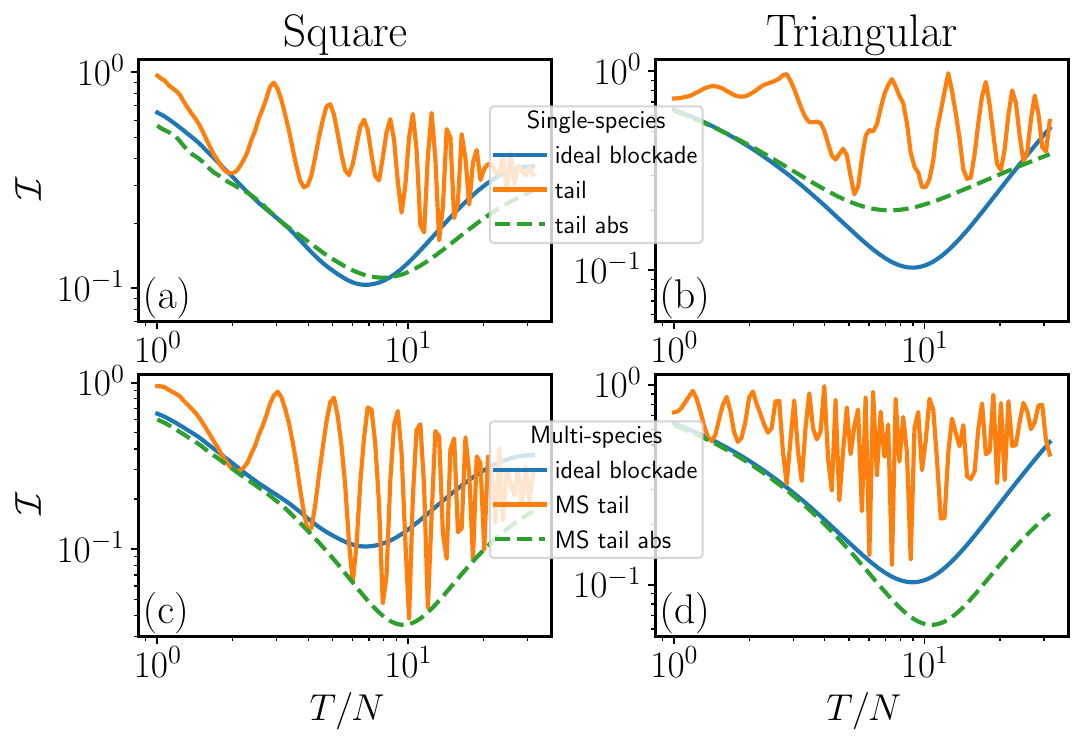}
    \caption{
        \textbf{Dynamical preparation with long-range tails.} Infidelities $\mathcal{I}$ obtained from periodic clusters with 6 unit cells and $\alpha=5$ as a function of $T/N$ for (a,b) single-species scenario and (c,d) multi-species scenario. 
        For square lattices, we set the final detuning $\Delta_f = 1.5$. In the single-species scenario, $r=0.16a$ and $R_{\mathrm{b}}=a \sqrt{1/4+(r/a)^{2}}$, whereas in the two-species scenario, $R^{\alpha\alpha}=\frac{a}{2}-r$ and $R^{\alpha\beta}=\sqrt{r^2+ \left( \frac{a}{2} \right)^2}$. 
        For triangular lattices, we set $\Delta_f = 3.0$. In the single-species scenario, $r=a/6$ and $R_{\mathrm{b}}=\frac{1}{2} \sqrt{  r ^{2} + \left( a + \sqrt{3} r \right)^2 }$, while in the three-species scenario, $r=0.11a$, $R^{\alpha\alpha}=\frac{a}{2}-r$, and $R^{\alpha\beta}=\frac{1}{2} \sqrt{r^2+\left(a+\sqrt{3}r\right)^2}$.} 
    \label{fig:App_tail}
\end{figure}

In the main text, we employ the blockade approximation, assuming infinitely large short-range interaction and negligible long-range interaction. However, in real experiments, interaction tails cannot be entirely ignored and will break the degeneracy of the dimer fully-packed states. 

Here, we demonstrate how the results of dynamical preparation change when tail effects are included. While still excluding states that violate the Rydberg blockade, we now account for the finite-range tail interactions beyond the blockade radius~\cite{Giuliano_PhysRevLett.129.090401}, represented by the Hamiltonian: $\hat{H}_{\mathrm{tail}}=\Omega_0 \sum_{ij} (\frac{R_{\rm b}}{|\bm{x}_{i}-\bm{x}_{j}|})^6 \hat{n}_{i} \hat{n}_{j}$. Here we consider the tails within the range $R_{\rm b}<|\bm{x}_{i}-\bm{x}_{j}|<2R_{\rm b}$.  The blockade radius, $R_{\rm b}$, is defined such that the interaction potential satisfies $V(|\bm{x}_{i}-\bm{x}_{j}|=R_{\rm b})=\Omega_0$, where $\Omega_0$ is the maximum Rabi frequency used during the dynamical preparation (cf.~Fig.~\ref{fig:DynamicalPreparation}(a)). 

Fig.~\ref{fig:App_tail} shows the overlap between the RVB state and the final state, as well as the absolute value of the final state, when tail effects are included. We consider both single-species and multi-species scenarios (cf.~S\ref{app:exp_imp}).  As discussed in Ref.~\cite{Giuliano_PhysRevLett.129.090401}, long-range tails break the energy degeneracy of the dimer fully-packed configurations, introducing phase differences among these configurations and leading to oscillations in the RVB overlap. When ignoring phase oscillations -- i.e. by considering the overlap between the RVB state and the absolute value of the final state -- we observe that the overlap remains of a comparable magnitude as in ideal blockade cases. 

Our results demonstrate that the presence of the tails does not significantly reduce the RVB overlap, and the multi-species scenario can effectively mitigate their impact. On square lattices, the single-species scenario yields an overlap comparable to that of the ideal blockade. The overlap can even surpass that of the ideal blockade when employing the two-species scenario. For triangular lattices, the overlap is generally worse compared to the ideal blockade (cf.~S\ref{app:cluster}), but using the three-species scenario can enhance the overlap to match the ideal blockade.

\section{Rydberg gadgets for the ``concentrated'' dimer constraint}
    
\begin{figure}[h]
    \centering  
    \includegraphics[width=0.55\columnwidth]{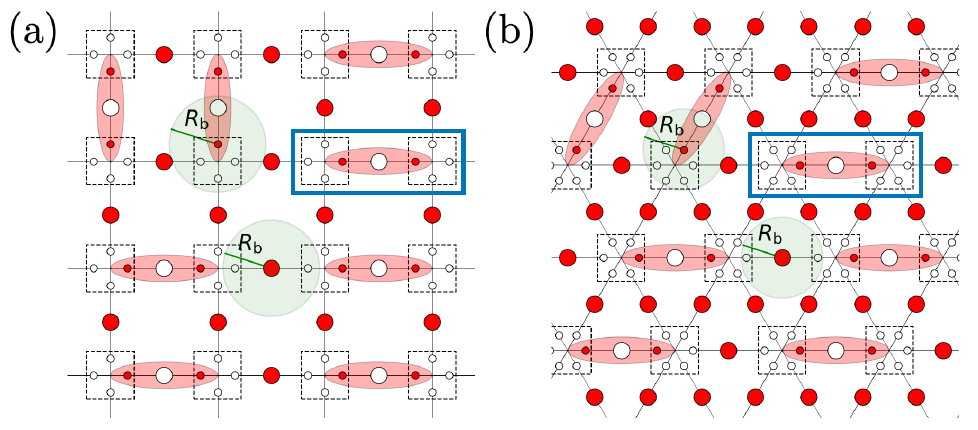}
    \caption{
        \textbf{Gadget method for constructing the concentrated dimer Hilbert space.}} 
    \label{fig:App_con}
\end{figure}

In this section, we discuss how to apply the gadget method to study another form of dimer constraint, namely what we call ``concentrated'' dimer constraint. This constraint is such that each vertex of the lattice is touched by \emph{at least} one dimer. 
Fig.~\ref{fig:App_con} illustrates the gadgets required to construct this constraint on square and triangular lattices. The atom arrangement is the same as with diluted dimer constraint, albeit with a shorter blockade radius. Specifically, each gadget atom is blockaded with all other gadget atoms on the same vertex and with the nearest edge atom. At the same time, each edge atom is blockaded with the two nearest gadget atoms on the corresponding vertices.
In this setup, a dimer is represented by the ground state of an edge atom, and it is accompanied by the nearest two Rydberg state gadget atoms. The low-energy subspace comprises minimally packed configurations in the $\Delta/\Omega \rightarrow \infty$ limit. The system permits the creation of additional dimers by de-exciting the edge atoms from the Rydberg state to the ground state.
Even though we do not perform large-scale numerical calculations for this model due to the prohibitive Hilbert space dimension compared with diluted models, we expect this model can also be used to dynamically prepare the QSL state in experiments.

\section{Rydberg gadgets for $\mathcal{Z}_2$ lattice gauge theory with dynamical matter}

\begin{figure}[h]
    \centering  
    \includegraphics[width=1.0\columnwidth]{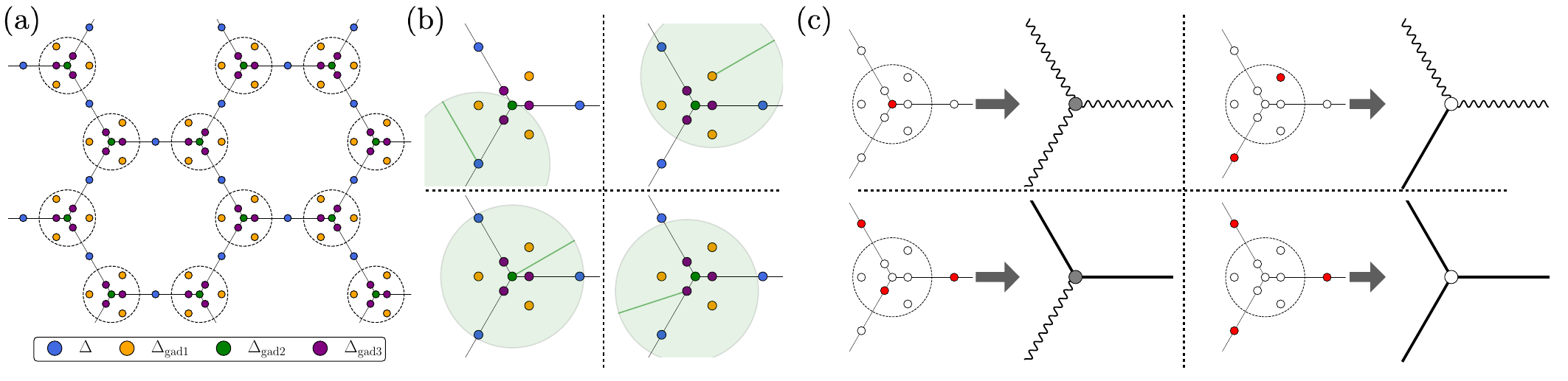}
    \caption{
        \textbf{Gadget method for constructing the constraint Hilbert space for $\mathcal{Z}_2$ LGT with dynamical matter.} (a) Schematic of the setup, where different colors indicate atoms with different detunings. (b) The $\mathcal{Z}_2$ LGT gadget and the blockade radius required to enforce the gauge constraint. (c) Correspondence between the four typical configurations in the gadget model and the LGT model. In the LGT model, the grey (white) dots represent the presence (absence) of matter, while wave (flat) lines represent $\hat{\tau}^x_{\langle i, j\rangle}=-1\ (+1)$ for the electric field.} 
    \label{fig:App_Z2LGT}
\end{figure}

In this section, we apply the gadget method to construct the constraint subspace for $\mathcal{Z}_2$ lattice gauge theory (LGT) with dynamical matter on honeycomb lattices \cite{PhysRevD.19.3682,Homeier2023}. Following the notation of Ref.~\cite{Homeier2023}, we consider the Hilbert space comprising matter (hard-core bosons $\hat{b}_{ j}$) on the vertices of lattices and the electric field (Pauli matrix $\hat{\tau}^x_{\langle i, j\rangle}$) on the links of lattices. 

The gauge symmetry (Gauss's law) requires the physical states to be eigenstates of the symmetry generators $\hat{G}_{j}= (-1)^{\hat{b}_{ j}^\dagger\hat{b}_{ j}}\prod_{ i:\langle i, j\rangle}\hat{\tau}^x_{\langle i, j\rangle}$ with eigenvalues $g_{ j}$ \cite{Homeier2023}. For the physical states ($g_{ j}=1$), at each vertex, if there exists (or does not exist) a matter such that $\hat{b}_{ j}^\dagger\hat{b}_{ j}=1$ (or $0$), the product of the electric fields associated with the vertex must be $-1$ (or $+1$). Quantum dimer models are special cases of these LGT models, which can be interpreted as LGT models with static matter.

In Fig.~\ref{fig:App_Z2LGT}, we illustrate the gadget method for constructing the constraint Hilbert space that satisfies Gauss's law.  The atoms are grouped into four types based on their detunings: the link, gadget $1$, gadget $2$, and gadget $3$ (cf.~Fig.~\ref{fig:App_Z2LGT}(a)). This can be realized with single-species atoms with local detuning control. Fig.~\ref{fig:App_Z2LGT}(b) demonstrates the isotropic blockade of these four types of atoms. Fig.~\ref{fig:App_Z2LGT}(c) illustrates the correspondence between four typical configurations in the gadget model and the LGT models.  In the gadget model, atoms on links in the Rydberg (ground) states represent the electric fields $\hat{\tau}^x_{\langle i, j\rangle}=+1\ (-1)$. Gadget atom $1$ in the Rydberg state, or all gadget atoms in the ground states represent the absence of matter ($\hat{b}_{ j}^\dagger\hat{b}_{ j}=0$), while gadget atoms $2$ or $3$ in the Rydberg states represent the presence of matter ($\hat{b}_{ j}^\dagger\hat{b}_{ j}=1$). This setup establishes a correspondence between the gadget model and the LGT model. In the classical limit ($\Omega=0$), by setting $ \Delta=\Delta_{\rm gad1} = 2\Delta_{\rm gad2}/3 = 2\Delta_{\rm gad3}$, these four configurations have the same energy (with an energy density per vertex of $3\Delta/2$), and form the system's lowest-energy subspace.

Compared to Ref.~\cite{Homeier2023}, our proposal can be implemented in a planar structure. In our scheme, the gauge constraint naturally emerges from the blockade mechanism, making the LGT subspace the system's lowest-energy subspace. However, while constructing the constraint Hilbert space is an essential first step toward simulating the $\mathcal{Z}_2$ LGT with dynamical matter, efficiently engineering the dynamics remains an open question and a promising direction for future study.

\end{document}